\documentclass[sigconf]{acmart}

\usepackage{booktabs} 
\usepackage{mathtools}
\usepackage{microtype}
\usepackage{listings}

\usepackage{lipsum}

\setcopyright{rightsretained}
\acmDOI{10.475/123_4}

\acmISBN{123-4567-24-567/08/06}

\acmConference[SPLASH'18]{ACM SIGPLAN International Conference on
Systems, Programming, Languages, and Applications: Software for
Humanity}{November 2018}{Boston, Massachusetts USA}
\acmYear{2018}
\copyrightyear{2018}

\acmArticle{4}
\acmPrice{15.00}

\newcommand{\sshom}{SSHOM}
\newcommand{\sshoms}{SSHOMs}

\begin{document}

\title{Finding Higher Order Mutants Using Variational Execution}
\titlenote{Accepted to SPLASH 2018 Student Research Competition}

\subtitle{Extended Abstract}

\author{Serena Chen}
\affiliation{%
  \institution{Franklin W. Olin College of Engineering}
  \streetaddress{1000 Olin Way}
  \city{Needham}
  \state{Massachusetts}
  \postcode{02492}
}


%
%

\maketitle


\section{Introduction}


Mutation testing is an effective but time consuming method for gauging the quality of a test suite. It functions by repeatedly making changes, called mutants, to the source code and checking whether the test suite fails (i.e., whether the mutant is \textit{killed})~\cite{demillo}. Recent work~\cite{jia2009higher} has shown cases in which applying multiple changes, called a \textit{higher order mutation}, is more difficult to kill than a single change, called a \textit{first order mutation}. Specifically, a special kind of higher order mutation, called a \textit{strongly subsuming higher order mutation (SSHOM)}, can enable equivalent accuracy in assessing the quality of the test suite with fewer executions of tests. Little is known about these \sshoms{}, as they are difficult to find. Jia et. al~\cite{jia2009higher} used a genetic search to find \sshoms; however, since it runs over many generations and has an element of randomness, the algorithm is time consuming and the results are often incomplete. 

Our goal in this research is to identify a faster, more reliable method for finding \sshoms{} in order to characterize them in the future. We propose an approach based on variational execution to find \sshoms. Preliminary results indicate that variational execution performs better than the existing genetic algorithm in terms of speed and completeness of results. Out of a set of 33 first order mutations, our variational execution approach finds all 38 \sshoms{} in 4.5 seconds, whereas the genetic algorithm only finds 36 of the 38 \sshoms{} in 50 seconds.

\section{Approach}


In this section, we discuss our approach to finding \sshoms{} using variational execution.

\subsection{Variational Execution}

Variational execution is a dynamic analysis that explores the behavior of a program under all inputs by sharing redundant executions. It was originally designed to examine the interactions of configuration options in a program~\cite{MWK+:ASE16}. The output of variational execution is an expression for the input conditions for which certain outputs are achieved.

We encode all possible first order mutants as runtime variables. An example is shown in Listing~\ref{lst:mut}. Then, we can use variational execution to efficiently and exhaustively explore the behavior of the test suite under all combinations of mutants. Variational execution will then output the mutant configurations in which a test case fails as a \textit{variability context}, or a propositional formula over the input options. For example, let's say we have two tests for the code example in Listing~\ref{lst:mut}:

\begin{enumerate}
    \item \texttt{assertTrue(equalsThree(1, 2));}
    \item \texttt{assertFalse(equalsThree(3, 4));}
\end{enumerate}

\noindent{}Variational execution would output two variability contexts:

\begin{enumerate}
    \item \texttt{\{test1 failed\} : (m0 $\wedge \neg$m1) $\vee$ ($\neg$m0 $\wedge$ m1)}
    \item \texttt{\{test2 failed\} : m1}
\end{enumerate}

\noindent{}From these variability contexts, we derive \sshoms{} using logical reasoning, which we will discuss next.

\definecolor{pgreen}{rgb}{0.25,0.5,0.37}
\newcommand{\lstbg}[3][0pt]{{\fboxsep#1\colorbox{#2}{\strut #3}}}
\lstdefinelanguage{diff}{
  basicstyle=\ttfamily,
  morecomment=[f][\lstbg{red!20}]-,
  morecomment=[f][\lstbg{green!20}]+,
}

\begin{lstlisting}[language=diff, basicstyle=\ttfamily, caption=Diff of mutations encoded as runtime variability., captionpos=b, frame=tb, float=tp, numbers=left, label=lst:mut, tabsize=2, escapechar=~]
boolean equalsThree(int a, int b) {
-  int c = a + b;
+  int c = m0 ? a - b : a + b;
-  return c == 3;
+  return m1 ? c != 3 : c == 3;
}
\end{lstlisting}

\subsection{Finding Strongly Subsuming Higher Order Mutants}

Current approaches to finding \sshoms{}, such as genetic algorithms or greedy algorithms, are search based~\cite{jia2009higher}. In contrast, our approach directly generates \sshoms{}. This approach uses the output of variational execution to construct a formula that encodes the definition of a \sshom{} and evaluates to true for all \sshoms{}. Since this formula is a satisfiability problem, we can use a BDD or SAT solver to cheaply find \sshoms{}.

To derive the formula, we outline the criteria for identifying \sshoms{} as defined by Jia et. al~\cite{jia2009higher} and construct a logical expression for each criterion.

Let $T$ be the set of all tests, $M$ be the set of all first order mutants, and $f(t)$ be the propositional formula for the mutant configurations in which test $t$ fails, as demonstrated in the variability contexts in the previous section. Let $\Gamma(m,t)$ be the result of evaluating $f(t)$ with first order mutant $m$; in other words, whether or not test $t$ fails with first order mutant $m$. The criteria for \sshoms{} follow:

\begin{enumerate}
    \item The \sshom{} must fail at least one test (must not be an equivalent mutant)
\end{enumerate}
\begin{equation}
\label{not-eq}
\bigvee\limits_{t \in T} f(t)
\end{equation}

 \noindent{}In order to find the mutant configurations in which at least one test fails, it is sufficient to take the disjunction of the conditions in which each test fails. Thus, if a mutant configuration successfully kills a test, the whole expression is true.

\begin{enumerate}
    \setcounter{enumi}{1}
    \item Every test that fails the \sshom{} must fail each constituent first order mutant
\end{enumerate}
\begin{equation}
\label{homsubset}
    \bigwedge\limits_{t \in T} (f(t) \Rightarrow \bigwedge\limits_{m \in M} {\underbrace{\vrule width0pt depth4pt ( \neg m \vee \Gamma(m, t) )}_{\strut\mathclap{ \text{Either } m \text{ is not included or is killed by }t}}})
\end{equation}

\noindent{}A certain mutant configuration (i.e. higher order mutant) killing a test implies one of two things for each first order mutant in the program: either the first order mutant is not in the higher order mutant ($\neg m$) or the test fails for the first order mutant ($\Gamma(m,t)$). This must hold over all tests and first order mutants. 

In addition, we can optimize for \sshoms{} that are harder to kill than the constituent first order mutants, excluding those that are equally difficult to kill~\cite{jia2009higher}. We call these \textit{strictly subsuming higher order mutants} and add a third criteria to find them. 

\begin{enumerate}
  \setcounter{enumi}{2}
  \item The set of tests that kill the higher order mutant must be a strict subset of the tests that kill all the first order mutants
\end{enumerate}
\begin{equation}
\label{strict-formula}
    \bigvee\limits_{t \in T} \big(\overbrace{\strut\neg f(t)}^{\mathclap{\strut\text{The selected mutants do not kill the test}}} \wedge \bigwedge\limits_{m \in M} {\underbrace{\vrule width0pt depth4pt ( \neg m \vee \Gamma(m, t) )}_{\strut\mathclap{ \text{Either } m \text{ is not included or is killed by }t}}}\big)
\end{equation}

\noindent{}If we replace all the constituent first order mutants with this strictly subsuming higher order mutant, then we will have a smaller set of tests that can kill the set of mutants. As such, there must exist at least one test that does not fail for the included mutants but does fail for each first order mutant of the higher order mutant.

To compute the \sshoms, we take the conjunction of Equations~\ref{not-eq}~and~\ref{homsubset} (and Equation~\ref{strict-formula} for strictly subsuming higher order mutants) and use a BDD or SAT solver to iteratively get all satisfiable solutions. Thus, this formula is guaranteed to generate all of the possible \sshoms{} for the given set of first order mutants, the given test set.

\section{Preliminary Evaluation}


We compared three approaches to finding \sshoms{}: a naive exhaustive search method, the state of the art genetic algorithm outlined in the work of Jia et. al~\cite{jia2009higher,harman2014angels}, and our variational execution approach.

The exhaustive search serves as a baseline. We iterate through all pairs of first order mutants, then all triples, and so on. For each candidate higher order mutant, we run the tests and check whether the results indicate that the higher order mutant is strongly subsuming. This algorithm has an exponential runtime.

As we could not find an existing implementation of the genetic algorithm, we reimplemented the algorithm presented by Jia et. al~\cite{jia2009higher,harman2014angels} with a few modifications optimizing for the current limited number of mutation operators.

We tested the three approaches on the program \textit{Triangle}. Triangle is a simple program that takes three integers representing three side lengths of a triangle and outputs whether the triangle is equilateral, scalene, isosceles, or invalid. This program is used in mutation testing literature and was used to test the original genetic algorithm for finding \sshoms~\cite{jia2009higher}. Triangle contains about 40 lines of code and 33 first order mutations. The test suite we used was generated by EvoSuite~\cite{Fraser:2011:EAT:2025113.2025179}, with a few test cases manually added for completeness.


\begin{figure}
    \centering
    \includegraphics[width=\linewidth]{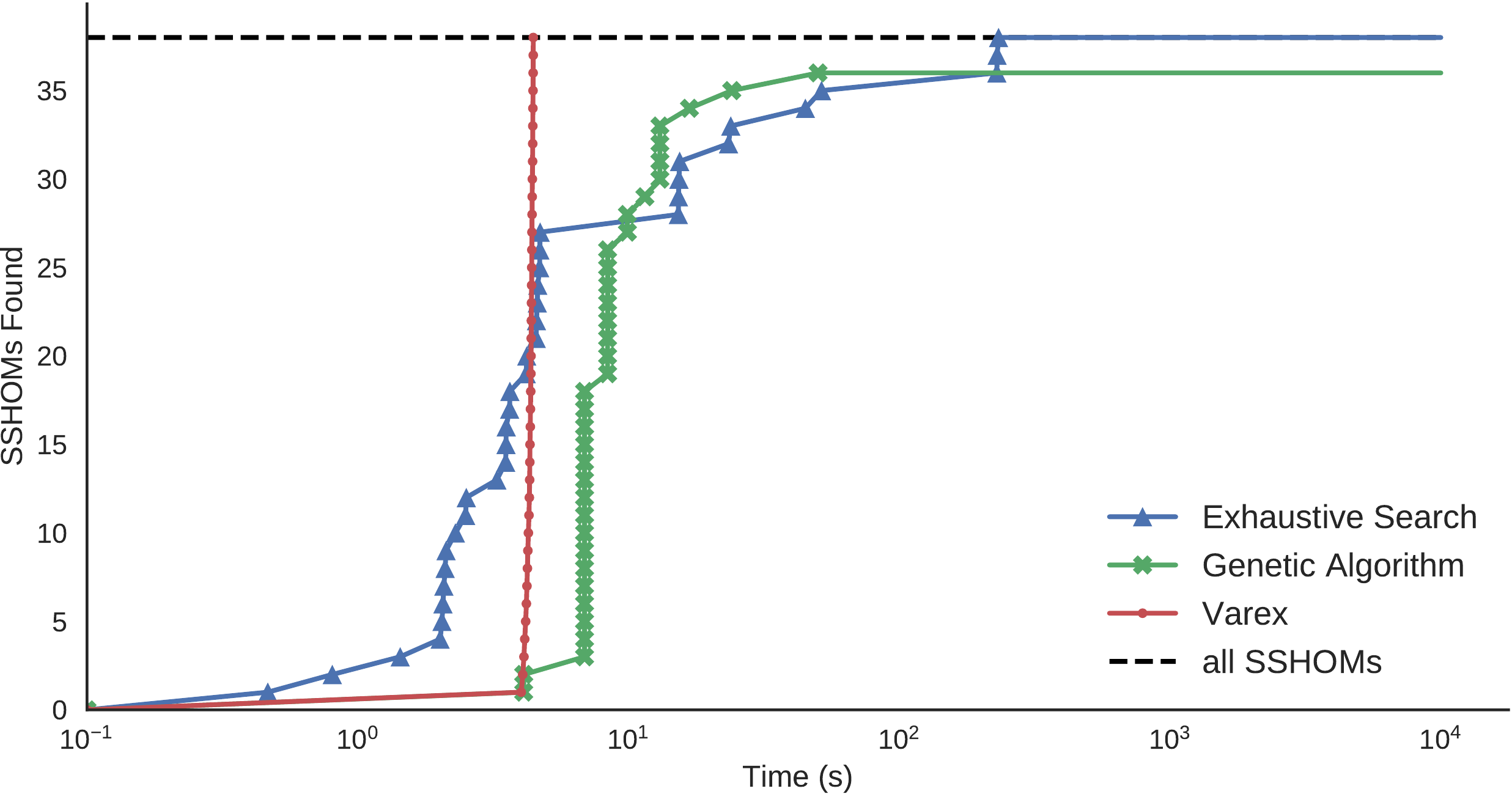}
    \caption{The number of mutants found over time in Triangle. Note, the x axis is in log scale.}
    \label{fig:comp-log}
\end{figure}

After running the three algorithms on Triangle, we plotted the number of \sshoms{} found over time, as shown in Figure \ref{fig:comp-log}. Since the variational execution and exhaustive search always find all the \sshoms{}, we establish that from the 33 first order mutations, there are 38 total \sshoms{}. The exhaustive search finds all the \sshoms{} in just under 4 minutes, but does not terminate as it continues testing more higher order mutants. The genetic algorithm finds its last \sshom{} after 50 seconds but it only finds 36 of the 38 \sshoms. Due to the random nature of the genetic algorithm, all four of our trials resulted in the genetic algorithm eventually hanging and no longer able to find more \sshoms. The variational execution approach finds all the \sshoms{} 50 times faster than the exhaustive search and 10 times faster than the genetic algorithm approach. The variational execution approach takes 4 seconds to run variational execution and 0.5 seconds to find all the \sshoms{}, a total of 4.5 seconds. Although variational execution takes a while, finding the \sshoms{} is very quick as satisfiability checks are inexpensive and fast compared to running a test suite, as happens in the exhaustive search and genetic algorithm. 

\section{Conclusions}


We propose a new way of finding \sshoms{} using variational execution and compared it to the state of the art genetic algorithm and a baseline exhaustive search method. The significant improvement in performance in a small program like Triangle leads us to believe that the variational execution approach is promising. In addition to extending to larger programs, future work includes analyzing the \sshoms{} and finding any defining characteristics.


\bibliographystyle{myabbrv_noaddress_nomonth}
\bibliography{bibliography}

\end{document}